\documentclass[a4paper,11pt]{article}

\usepackage[utf8]{inputenc}
\usepackage[top=50pt,bottom=80pt,left=68pt,right=66pt]{geometry}
\usepackage{amsmath, amssymb}
\usepackage{mathrsfs}
\usepackage{cite}
\usepackage{float}
\usepackage{color}
\usepackage{xcolor}
\usepackage{multirow}
\usepackage{graphicx,caption,subcaption}
\usepackage{bm}
\usepackage{braket}
\usepackage{dsfont}
\usepackage{cancel}
\usepackage{latexsym}
\usepackage{slashed}
\usepackage[space]{grffile}

\definecolor{darkblue}{rgb}{0.1,0.1,.8}
\usepackage[breaklinks=true,linktocpage=true,
colorlinks=true,urlcolor=darkblue,linkcolor=darkblue,
citecolor=darkblue,pdfpagelabels=true,hypertexnames=true,
plainpages=false,naturalnames=false,]{hyperref}

\numberwithin{equation}{section}

\begin{document}

\renewcommand{\baselinestretch}{1.2}
\thispagestyle{empty}

\mbox{}

\vskip3.0cm
\begin{center}

{\LARGE\bf Bispectrum at NLO in single field inflation: conservation and squeezed limit}

\vglue.3in

\hspace{-2pt}Ignatios Antoniadis,${}^{a,b,}$\footnote{antoniad@lpthe.jussieu.fr} Jules Cunat,${}^{b,}$\footnote{jcunat@lpthe.jussieu.fr} Auttakit Chatrabhuti,${}^{a,}$\footnote{auttakit.c@chula.ac.th} Hiroshi Isono${}^{a,}$\footnote{hiroshi.i@chula.ac.th}
\vglue.1in

${}^a$~{\it Department of Physics, Faculty of Science, Chulalongkorn University,\\ Phayathai Road, Pathumwan, Bangkok 10330, Thailand}\\
${}^b$~{\it Laboratoire de Physique Th\'eorique et Hautes Energies (LPTHE), Sorbonne Universit\'e,\\ CNRS, 4 Place Jussieu, 75005 Paris, France}
\vglue.1in

\end{center}


%
%
%

\vglue.3in

\begin{center}
{\Large\bf Abstract}
\vglue.2in
\end{center}

In a recent paper, we computed the bispectrum of primordial density perturbations in CMB to second order in the slow-roll parameters of single field inflation, and found logarithmic infrared contributions that diverge in both large physical distances and squeezed limit where one momentum vanishes. In this work, we provide an independent test of the result by checking its conservation and the validity of the consistency relation between the squeezed limit of the bispectrum and the square of the power spectrum. Despite the violation of the main assumption for its general proofs which is the finiteness of the relevant observables in these limits, we find that the identity continues to hold in the vicinity of the squeezed limit and large time.

\newpage

\tableofcontents

\setcounter{footnote}{0}

\section{Introduction}
It is well known that propagation of massless fields in de Sitter (dS) space is plagued with infrared divergences at distances larger than than the dS radius $H^{-1}$, with $H$ the Hubble constant fixing the expansion rate. In particular, the propagator of a massless minimally coupled scalar, like the inflaton during inflation, grows logarithmically~\cite{Ford:1977in,Ratra:1984yq,Antoniadis:1985pj}. These divergences are physical and show up in inflationary observables, as we found recently in the computation of the three-point function of scalar primordial fluctuations (bispectrum) at the next-to-leading order (NLO) in the slow-roll parameters of single field inflation, giving rise to sizeable corrections in the non-gaussianity parametrised by an observable parameter $f_\text{NL}$. 
In summary, we found two types of logarithmic corrections to the bispectrum~\cite{Antoniadis:2024abm}: 
\begin{enumerate}
\item one is proportional to $\ln(-k_\text{t}\tau)$ where $k_\text{t}$ is the sum of the three comoving momenta $k_\text{t}=\sum_{i=1}^3k_i$ with $k_i=|\vec{k}_i|$ and $\tau$ is the conformal time that vanishes at large proper time $t$, $\tau H\simeq-e^{Ht}$; 
\item the other is proportional to $\ln(k_i/k_\text{t})$ which is time independent but diverges when one of the momenta vanishes. 
\end{enumerate}
The first correction diverges at large physical distances but is finite in the squeezed limit where one of the momenta vanishes, while the second diverges in the squeezed limit. 

In practice, the first correction is the largest since physical distances probed in CMB anisotropies are of order $0.01-100$ Mpc $\sim 10^{20}-10^{24}$m and using a typical inflation scale $\sim 10^{12}$ GeV corresponding to a dS radius $\sim 10^{-28}$m, one finds that the value of the logarithm is at least of order 100 and compensates the NLO suppression of one slow-roll parameter. On the other hand, the argument of the second logarithm is the ratio of the minimum to the maximum values of momenta probed in CMB observations $k_\text{min}/k_\text{max}$, which is about four orders of magnitude, corresponding to the physical distances mentioned above.

Theoretically, however, these divergences invalidate the assumption of finiteness of the relevant (appropriately normalised) quantities entering in the formal proofs that relate the squeezed limit of the bispectrum to the square of the power spectrum that we recall here.
The power spectrum\footnote{Note that the power spectrum is often defined as $\frac{k^3}{2\pi^2}\mathcal{P}$.} and the bispectrum are defined as
\begin{align}
\big\langle\zeta\big(\vec{k}_1\big)\zeta\big(\vec{k}_2\big)\big\rangle&=(2\pi)^3\delta^{(3)}\big(\vec{k}_1+\vec{k}_2\big)\mathcal{P}_{k_1}, \\
    \big\langle\zeta\big(\vec{k}_1\big)\zeta\big(\vec{k}_2\big)\zeta\big(\vec{k}_3\big)\big\rangle&=(2\pi)^3\delta^{(3)}\big(\vec{k}_1+\vec{k}_2+\vec{k}_3\big)\mathcal{B}(k_1,k_2,k_3),
\end{align}
where $\zeta$ is the gauge invariant scalar perturbation.
The squeezed limit of the bispectrum is given by the well-known consistency relation \cite{Maldacena:2002vr,Creminelli:2004yq}
\begin{align}\label{eq:consistencyrelation} 
&\big\langle\zeta\big(\vec{k}_1\big)\zeta\big(\vec{k}_2\big)\zeta\big(\vec{k}_3\big)\big\rangle_{k_1\rightarrow0}
\sim -(2\pi)^3\delta^{(3)}\big(\vec{k}_1+\vec{k}_2+\vec{k}_3\big)
\mathcal{P}_{k_1}\mathcal{P}_{k_2}\frac{d\ln(k_2^3\mathcal{P}_{k_2})}{d\ln(k_2)}
\left[1+\mathcal{O}\left(\frac{k_1}{k_2}\right)\right] \nonumber\\
&\qquad\qquad\qquad\quad\sim 
-(2\pi)^3\delta^{(3)}\big(\vec{k}_1+\vec{k}_2+\vec{k}_3\big)
\mathcal{P}_{k_1}\left(3+k_2\frac{d}{dk_2}\right)\mathcal{P}_{k_2}
\left[1+\mathcal{O}\left(\frac{k_1}{k_2}\right)\right],
\end{align}
or, by introducing the scalar tilt $n_s$ as $\mathcal{P}_k\sim k^{n_s-3}$:
\begin{align}
    \big\langle\zeta\big(\vec{k}_1\big)\zeta\big(\vec{k}_2\big)\zeta\big(\vec{k}_3\big)\big\rangle_{k_1\rightarrow0}\sim -n_s(2\pi)^3\delta^{(3)}\big(\vec{k}_1+\vec{k}_2+\vec{k}_3\big)\mathcal{P}_{k_1}\mathcal{P}_{k_1}
\left[1+\mathcal{O}\left(\frac{k_1}{k_2}\right)\right].
\end{align}

Note that all quantities above are assumed to be time independent (existence of large time limit) and the behaviour in the limit $k_1\rightarrow0$ of the 3-point function to be the same as the behaviour of $\mathcal{P}_{k_1}$, so that the limit exists upon multiplication with $k_1^{3-n_s}$ ($k_1^3$ at lowest order), or equivalently $\lim_{k_1\to0}(\mathcal{B}_{k_1,k_2,k_3}/\mathcal{P}_{k_1})$ is finite. 
Obviously, the existence of logarithmic divergences described above invalidates these assumptions and asks the question on the validity of the consistency relation in the squeezed limit, or its eventual modification when these divergences are taken into account.
In this note, we study this question and we find that the consistency relation remains valid in the vicinity of these limits, at least to NLO where explicit expressions are known. We also show that the NLO expression of the bispectrum~\cite{Antoniadis:2024abm} is conserved in time.

Actually the squeezed limit of the bispectrum and the consistency relation up to NLO were studied in~\cite{Renaux-Petel:2010paw} using a different method based on taking the limit in the action, rather than directly of the correlator. Our analysis serves also as an independent check of our previous general result for the bispectrum at NLO.

The outline of our paper is the following. In Section~2, we recall the derivation of the consistency relation based on a global scaling symmetry. In Section~3, we present the bispectrum up to NLO using the results of our previous work and discuss its time independence. We also discuss the time independence of the non-gaussianity parameter $f_\text{NL}$ and justify a conserved (time independent) definition beyond leading order.
In Section~4, we obtain the squeezed limit of the bispectrum and the power spectrum and verify the consistency relation as well as that for different horizon exit times. Finally, Section~5 contains our conclusions.

Before proceeding, we summarise our conventions. We consider the background inflationary metric to be $ds^2=-dt^2+a(t)^2d\vec x^2$ with $\vec x=(x^1,x^2,x^3)$. The Hubble parameter is defined by $H=\dot a/a$, where the dot is $d/dt$, and the slow-roll parameters are defined by $\varepsilon_1=-\dot H/H^2$ and $\varepsilon_{i+1}=\dot\varepsilon_i/H\varepsilon_i$ ($i\geq 1$)\footnote{In~\cite{Antoniadis:2024abm}, we called $\varepsilon_i$ the Hubble flow functions in order to emphasise the difference from the slow-roll parameters defined through an inflaton potential that will not appear in this paper.}. Note that these \emph{parameters} are functions of time. The conformal time $\tau$ is defined through $d\tau=dt/a(t)$.

\section{The consistency relation}
We start by briefly deriving the consistency relation \cite{Creminelli:2004yq,Cheung:2007sv,Creminelli:2011rh,Creminelli:2012ed}\footnote{See also~\cite{Goldberger:2013rsa,Pimentel:2013gza,Kundu:2015xta,Hui:2018cag} for other derivations of the consistency relation and its generalisations.}. In the case where one of the three momenta is much smaller than the others, say $k_1\ll k_2,k_3$, it exits the horizon way before. Consequently, it is a frozen classical mode acting as the background for the other two. In order to compute the 3-point function in this limit, one can therefore multiply $\zeta\big(\vec{k}_1\big)$ by the 2-point function of $\zeta\big(\vec{k}_2\big)$ and $\zeta\big(\vec{k}_3\big)$ with $\zeta\big(\vec{k}_1\big)$ acting as a background and average over it. Namely we can write
\begin{align}\label{eq:withbackground}
    \big\langle\zeta\big(\vec{k}_1\big)\zeta\big(\vec{k}_2\big)\zeta\big(\vec{k}_3\big)\big\rangle_{k_1\rightarrow\,0}\sim\Big\langle\zeta_L\big(\vec{k}_1\big)\big\langle\zeta\big(\vec{k}_2\big)\zeta\big(\vec{k}_3\big)|\zeta_L\big\rangle\Big\rangle_{k_1\rightarrow\,0}.
\end{align}
One can then compute the 2-point function of $\zeta\big(\vec{k}_2\big)$ and $\zeta\big(\vec{k}_3\big)$ in this new background $\zeta_L$ by doing a perturbative expansion around the true background, i.e. without the long wavelength mode $\zeta_L$ which is almost constant in position space and thus has a sharp peak around $\vec k_1=0$ in momentum space. 

Actually, one can use a trick here. One can show that  the long wavelength mode $\zeta_L\big(\vec{k}_1\big)$ can be seen as an adiabatic mode which can be added or subtracted by a change of coordinates \cite{Creminelli:2012ed,Weinberg:2003sw}. At leading order in $k_1$, it can be removed by a simple time-independent spatial dilatation. To see this, let us recall that the gauge has already been fixed here in the $\zeta$-gauge where the spatial distance is given by
\begin{align}
    d\vec{\ell}\,^2=a^2(t)e^{2\zeta(t,\vec{x})}d\vec{x}^2.
\end{align}
So performing a time-independent spatial dilatation\footnote{Indeed the long wavelength mode can be considered as frozen. Its time dependence would be subleading in $k_L$. Moreover, we could also consider a special conformal transformation but its effect is also subleading. See \cite{Creminelli:2012ed} for more details.}
\begin{align}
    \vec{x}\rightarrow e^{\zeta_L}\vec{x},
\end{align}
we do not change the gauge but introduce a long wavelength mode as the spatial line element is now
\begin{align}
    d\vec{\ell}\,^2\rightarrow a^2e^{2\zeta(t,\vec{x})+2\zeta_L}d\vec{x}^2.
\end{align}
Consequently, in position space, the 2-point function of $\zeta\big(\vec{k}_2\big)$ and $\zeta\big(\vec{k}_3\big)$ in the background of the long wavelength mode $\zeta_L\big(\vec{k}_1\big)$ can be written as
\begin{align}
    \big\langle\zeta(\vec{x}_2)\zeta(\vec{x}_3)|\zeta_L\big\rangle&=\big\langle\zeta(\vec{\tilde{x}}_2)\zeta(\vec{\tilde{x}}_3\big)\big\rangle \nonumber\\
    &=\big\langle\zeta(\vec{x}_2)\zeta(\vec{x}_3\big)\big\rangle+\zeta_L(x_B)\Big(\vec{x}_2\cdot\vec{\partial}_{\vec{x}_2}+\vec{x}_{3}\cdot\vec{\partial}_{\vec{x}_3}\Big)\big\langle\zeta(\vec{x}_2)\zeta(\vec{x}_3\big)\big\rangle+\cdots,
    \label{eq:backgroundexpansion}
\end{align}
where we have performed the infinitesimal change of coordinates
\begin{align}
    \vec{x}\rightarrow\vec{\tilde{x}}=\big(1+\zeta_L\big)\vec{x}.
\end{align}

The first term of \eqref{eq:backgroundexpansion} will give a vanishing contribution when one re-introduces it in \eqref{eq:withbackground} and averages over the long wavelength mode. Moreover, its position $\vec x_B$ was chosen so that $\zeta(\vec x_B)$ provides a representative value of the long wavelength (namely almost constant) $\zeta_L$. Its choice is arbitrary 
and we are now interested in the second term of \eqref{eq:backgroundexpansion}:
\begin{align}\label{1ststep}
\begin{split}
    &\quad\zeta_L(x_B)\Big(\vec{x}_2\cdot\vec{\partial}_{\vec{x}_2}+\vec{x}_3\cdot\vec{\partial}_{\vec{x}_3}\Big)\big\langle\zeta(x_2)\zeta(x_3\big)\big\rangle \\
    &=\int\frac{d^3\vec{k}_B}{(2\pi)^3}\frac{d^3\vec{k}_2}{(2\pi)^3}\frac{d^3\vec{k}_3}{(2\pi)^3}(2\pi)^3\delta^{(3)}\big(\vec{k}_2+\vec{k}_3\big)\zeta_L\big(\vec{k}_B\big)\mathcal{P}\big(\vec{k}_2,\vec{k}_3\big) \\
    &\qquad\times\Big(\vec{k}_2\cdot\vec{\partial}_{\vec{k}_2}+\vec{k}_3\cdot\vec{\partial}_{\vec{k}_3}\Big)e^{i\vec{x}_B\cdot\vec{k}_B+i\vec{x}_2\cdot\vec{k}_2+i\vec{x}_3\cdot\vec{k}_3} \\
    &=-\int\frac{d^3\vec{k}_B}{(2\pi)^3}\frac{d^3\vec{k}_2}{(2\pi)^3}\frac{d^3\vec{k}_3}{(2\pi)^3}(2\pi)^3\zeta_L\big(\vec{k}_B\big)e^{i\vec{x}_B\cdot\vec{k}_B+i\vec{x}_2\cdot\vec{k}_2+i\vec{x}_3\cdot\vec{k}_3} \\
    &\qquad\quad\times\Big[6\delta^{(3)}\big(\vec{k}_2+\vec{k}_3\big)\mathcal{P}\big(\vec{k}_2,\vec{k}_3\big)+\delta^{(3)}\big(\vec{k}_2+\vec{k}_3\big)\Big(\vec{k}_2\cdot\vec{\partial}_{\vec{k}_2}+\vec{k}_3\cdot\vec{\partial}_{\vec{k}_3}\Big)\mathcal{P}\big(\vec{k}_2,\vec{k}_3\big) \\
    &\qquad\qquad\quad+\mathcal{P}\big(\vec{k}_2,\vec{k}_3\big)\big(\vec{k}_2+\vec{k}_3\big)\cdot\vec{\partial}_{\vec{k}_2+\vec{k}_3}\delta^{(3)}\big(\vec{k}_2+\vec{k}_3\big)\Big] \\
    &=-\int\frac{d^3\vec{k}_B}{(2\pi)^3}\frac{d^3\vec{k}_2}{(2\pi)^3}\frac{d^3\vec{k}_3}{(2\pi)^3}(2\pi)^3\zeta_L\big(\vec{k}_B\big)e^{i\vec{x}_B\cdot\vec{k}_B+i\vec{x}_2\cdot\vec{k}_2+i\vec{x}_3\cdot\vec{k}_3} \\
    &\quad\qquad\times\Big[3\delta^{(3)}\big(\vec{k}_2+\vec{k}_3\big)\mathcal{P}\big(\vec{k}_2,\vec{k}_3\big)+\delta^{(3)}\big(\vec{k}_2+\vec{k}_3\big)\Big(\vec{k}_2\cdot\vec{\partial}_{\vec{k}_2}+\vec{k}_3\cdot\vec{\partial}_{\vec{k}_3}\Big)\mathcal{P}\big(\vec{k}_2,\vec{k}_3\big)+\cdots\Big],
\end{split}
\end{align}
where the integration over $\vec{k}_B$ is around the origin $\vec k_B=0$ where $\zeta_L$ peaks in momentum space. In \eqref{1ststep} we performed two integrations by part, and the dots in the last line denote vanishing terms as they are proportional to $(\vec{k}_2+\vec{k}_3)\delta^{(3)}(\vec{k}_2+\vec{k}_3)$. Moreover, it is important to notice that $\mathcal{P}(\vec{k}_2,\vec{k}_3)$ is the power spectrum without the momentum conservation $\vec k_3=-\vec k_2$ from $\delta^{(3)}(\vec{k}_2+\vec{k}_3)$ applied. It is related to $\mathcal{P}_{k_2}$ as
\begin{align}
    \mathcal{P}_{k_2}=\mathcal{P}\big(\vec{k}_2,-\vec{k}_2\big)
\end{align}
and therefore 
\begin{align}
    \Big(\vec{k}_2\cdot\vec{\partial}_{\vec{k}_2}+\vec{k}_3\cdot\vec{\partial}_{\vec{k}_3}\Big)\mathcal{P}\big(\vec{k}_2,\vec{k}_3\big)=k_2\frac{d}{dk_2}\mathcal{P}_{k_2}
\end{align}
under the delta function.

The last step is now to multiply \eqref{eq:backgroundexpansion} by $\zeta_L\big(\vec{k}_1\big)$ and to average over it. We are then left over only with the second term on the right hand side which was evaluated in \eqref{1ststep}. In Fourier space and after integrating over the momentum $\vec{k}_B$ we then obtain
\begin{align}
\label{zetaL}
    \Big\langle\zeta_L\big(\vec{x}_1\big)\big\langle\zeta\big(\vec{x}_2\big)\zeta\big(\vec{x}_3\big)|\zeta_L\big\rangle\Big\rangle
    &=-\int\frac{d^3\vec{k}_1}{(2\pi)^3}\frac{d^3\vec{k}_2}{(2\pi)^3}\frac{d^3\vec{k}_3}{(2\pi)^3}e^{i\vec{x}_1\cdot\vec{k}_1+i\vec{x}_2\cdot\vec{k}_2+i\vec{x}_3\cdot\vec{k}_3}(2\pi)^3\delta^{(3)}\big(\vec{k}_2+\vec{k}_3\big) \nonumber\\
    &\qquad\qquad\qquad\times
\mathcal{P}_{k_1}\left(3\mathcal{P}_{k_2}+k_2\frac{d\mathcal{P}_{k_2}}{dk_2}\right)
\left[1+\mathcal{O}\left(\frac{k_1}{k_2}\right)\right],
\end{align}
where we used the fact that $\vec{k}_1\sim 0$ while $\vec{x}_B$ is arbitrary and can be chosen to be around the origin.
On the other hand, by definition, we also have
\begin{align}
    \begin{split}
\big\langle\zeta\big(\vec{x}_1\big)\zeta\big(\vec{x}_2\big)\zeta\big(\vec{x}_3\big)\big\rangle
&=\int\frac{d^3\vec{k}_1}{(2\pi)^3}\frac{d^3\vec{k}_2}{(2\pi)^3}\frac{d^3\vec{k}_3}{(2\pi)^3}
e^{i\vec{x}_1\cdot\vec{k}_1+i\vec{x}_2\cdot\vec{k}_2+i\vec{x}_3\cdot\vec{k}_3} \nonumber\\
&\qquad\qquad\qquad \times(2\pi)^3\delta^{(3)}\big(\vec{k}_1+\vec{k}_2+\vec{k}_3\big)\mathcal{B}(k_1,k_2,k_3).
    \end{split}
\end{align}
Equating this with \eqref{zetaL} as claimed by \eqref{eq:withbackground}, we obtain
\begin{align}\label{eq:consistencyrelation2}
\mathcal{B}(k_1,k_2,k_3)_{k_1\rightarrow0}=-\mathcal{P}_{k_1}
\left(3\mathcal{P}_{k_2}+k_2\frac{d\mathcal{P}_{k_2}}{dk_2}\right)
\left[1+\mathcal{O}\left(\frac{k_1}{k_2}\right)\right],
\end{align}
which is indeed \eqref{eq:consistencyrelation}. 

We recall that one assumption in the above derivation is that the inflaton is the only degree of freedom, so that the long wavelength modes are adiabatic and can be removed by a change of coordinates. An additional assumption however was implicit in the above arguments, namely the finiteness of the 2-point and 3-point functions at large physical distances, as well as of the ratio $\mathcal{B}_{k_1,k_2,k_3}/\mathcal{P}_{k_1}$ as $k_1\to 0$. These may be invalidated by the presence of infrared divergences and an explicit verification is needed, which we perform here up to NLO in slow-roll parameters.

\section{Bispectrum up to NLO}

\subsection{Results and time independence}

Here, we review the bispectrum up to NLO~\cite{Antoniadis:2024abm} in single field inflation and discuss its time independence. We first introduce the amplitude $\mathscr{A}$ of the bispectrum $\mathcal{B}$ through the Hubble and slow-roll parameters at the time $\tau$ when the 3-point function is defined:
\begin{align}\label{eq:BandAtau}
\mathcal{B}(k_1,k_2,k_3)=\frac{H^{4}}{32\varepsilon_1^{2}}\frac{\mathscr{A}}{k_1^3k_2^3k_3^3}. 
\end{align}
The amplitude $\mathscr{A}$ is then given by
\begin{align}
\mathscr{A}&=\mathscr{A}_{\text{LO}}+\mathscr{A}_{\text{NLO}}+\mathcal{O}(\varepsilon^3), \label{bispectrumtau}
\end{align}
where $\mathscr{A}_{\text{LO}}$ and $\mathscr{A}_{\text{NLO}}$ are the first and second order parts of $\mathscr{A}$ in terms of the slow-roll parameters, respectively:
\begin{align}
\mathscr{A}_{\text{LO}}&=\varepsilon_1K_1\big(k_1,k_2,k_3\big)+\varepsilon_2K_2\big(k_1,k_2,k_3\big)+\mathcal{O}(\tau), 
\label{ALO}\\
\begin{split}
    \mathscr{A}_{\text{NLO}}&=\left[\gamma_E-1+\ln(\!-\tau k_\text{t})\right]\Big(\varepsilon_1^2K_3\big(k_1,k_2,k_3\big)\\
    &\qquad\quad+\varepsilon_1\varepsilon_2K_4\big(k_1,k_2,k_3\big)+\varepsilon_2^2K_5\big(k_1,k_2,k_3\big)+\varepsilon_2\varepsilon_3K_6\big(k_1,k_2,k_3\big)\Big)\\
    &\quad~+\big(2\varepsilon_1+\varepsilon_2\big)\ln\left(\frac{2k_1}{k_\text{t}}\right)\Big(\varepsilon_1K_7\big(k_1,k_2,k_3\big)+\varepsilon_2K_8\big(k_1,k_2,k_3\big)\Big)\\
    &\quad~+\big(2\varepsilon_1+\varepsilon_2\big)\ln\left(\frac{2k_2}{k_\text{t}}\right)\Big(\varepsilon_1K_7\big(k_2,k_1,k_3\big)+\varepsilon_2K_8\big(k_2,k_1,k_3\big)\Big)\\
    &\quad~+\big(2\varepsilon_1+\varepsilon_2\big)\ln\left(\frac{2k_3}{k_\text{t}}\right)\Big(\varepsilon_1K_7\big(k_3,k_2,k_1\big)+\varepsilon_2K_8\big(k_3,k_2,k_1\big)\Big)\\
    &\quad~+\varepsilon_1^2K_9\big(k_1,k_2,k_3\big)+\varepsilon_1\varepsilon_2K_{10}\big(k_1,k_2,k_3\big)\\
    &\quad~+\varepsilon_2^2K_{11}\big(k_1,k_2,k_3\big)+\varepsilon_2\varepsilon_3K_{12}\big(k_1,k_2,k_3\big)+\mathcal{O}(\tau),\end{split} \label{ANLO}
\end{align}
where the functions $K_i$ are given by:
\begin{align}
    \begin{split}
    K_1\big(k_1,k_2,k_3\big)&=8\frac{k_1^2k_2^2+k_1^2k_3^2+k_2^2k_3^2}{k_\text{t}}-2(k_1^3+k_2^3+k_3^3)
    +k_\text{t}(k_1^2+k_2^2+k_3^2), \\
    K_2\big(k_1,k_2,k_3\big)&=k_1^3+k_2^3+k_3^3, \\
    K_3\big(k_1,k_2,k_3\big)&=-4K_1\big(k_1,k_2,k_3\big), \\
    K_4\big(k_1,k_2,k_3\big)&=-K_1\big(k_1,k_2,k_3\big)-4K_2\big(k_1,k_2,k_3\big), \\
    K_5\big(k_1,k_2,k_3\big)&=-2K_2\big(k_1,k_2,k_3\big), \\
    K_6\big(k_1,k_2,k_3\big)&=K_2\big(k_1,k_2,k_3\big), \\
    K_7\big(k_1,k_2,k_3\big)&=\frac{k_2+k_3}{-k_1+k_2+k_3}\Big(2k_1k_2k_3-K_1\big(k_1,k_2,k_3\big)\Big), \\
    K_8\big(k_1,k_2,k_3\big)&=k_1^3-K_2\big(k_1,k_2,k_3\big), \\
    K_9\big(k_1,k_2,k_3\big)&=2k_1k_2k_3-K_1\big(k_1,k_2,k_3\big),\\
    K_{10}\big(k_1,k_2,k_3\big)&=2K_1\big(k_1,k_2,k_3\big)-3K_2\big(k_1,k_2,k_3\big)-3k_1k_2k_3 +k_\text{t}(k_1^2+k_2^2+k_3^2), \\
    K_{11}\big(k_1,k_2,k_3\big)&=2K_2\big(k_1,k_2,k_3\big),\\
    K_{12}\big(k_1,k_2,k_3\big)&=k_1k_2k_3+K_2\big(k_1,k_2,k_3\big) 
    -k_\text{t}(k_1^2+k_2^2+k_3^2).
    \end{split}
\label{Ks}
\end{align}

Let us consider the time (in)dependence of the bispectrum. In general, the computation of higher order corrections in slow-roll parameters involves the expansion of the Hubble and slow-roll parameters in terms of their small time derivatives (namely, the slow-roll parameters) around a reference time, which we call a pivot time. It can be chosen arbitrarily. For example, in the computation of the bispectrum in~\cite{Antoniadis:2024abm}, the pivot time was chosen to be the time of the correlator $\tau$, so that the Hubble and slow-roll parameters in $\mathcal{B}$ above are evaluated at $\tau$. Therefore, the dependence of $\mathcal{B}$ on $\tau$ comes from not only the logarithmic term $\ln(-k_{\text{t}}\tau)$ but also the parameters $H,\varepsilon_i$. A natural question is then how the expression of the bispectrum $\mathcal{B}$ changes when $H,\varepsilon_i$ are expanded around another pivot time $\tau^*$. After the expansion, the bispectrum is expressed in $H,\varepsilon_i$ evaluated at $\tau^*$ and will contain new powers of $\ln(\tau/\tau^*)$ which come from the expansion\footnote{This logarithm is nothing but $H(t_*-t)$ in terms of the proper time at leading order in the slow-roll parameters.}. It should be noted that this is just rewriting the bispectrum, manifesting the $\tau$ dependence of $H,\varepsilon_i$ through their expansion. Now, if the bispectrum after the expansion around $\tau^*$ turns out to be independent of $\tau$, one may claim that the bispectrum, which is defined at time $\tau$, is actually independent of $\tau$. To put it simply, this means that the $\tau$ dependence of $H,\varepsilon_i$ compensates the pre-existing logarithm $\ln(-\tau)$.

Let us demonsterate this. We first introduce the leading order (LO) part $\mathcal{B}_{\text{LO}}$ and the NLO part $\mathcal{B}_{\text{NLO}}$ by
\begin{align}
&\mathcal{B}_{\text{LO}}=\frac{H^4}{32\varepsilon_1^2}\frac{\mathscr{A}_{\text{LO}}}{k_1^3k_2^3k_3^3}, \quad
\mathcal{B}_{\text{NLO}}=\frac{H^4}{32\varepsilon_1^2}\frac{\mathscr{A}_{\text{NLO}}}{k_1^3k_2^3k_3^3},
\end{align}
where $H,\varepsilon_i$ are evaluated at $\tau$.
We now expand them around another pivot time $\tau^*$.
For the LO part $\mathcal{B}_{\text{LO}}$, it is enough to expand $H$ and $\varepsilon_i$ up to the next order. The expansions involve logarithmic factors:
\begin{align}
H(t(\tau))&=H^*\left[1+\varepsilon_1^*\ln\left(\frac{\tau}{\tau^*}\right)+\mathcal{O}(\varepsilon^{*2})\right], \\
\varepsilon_i(t(\tau))&=\varepsilon_i^*\left[1-\varepsilon_{i+1}^*\ln\left(\frac{\tau}{\tau^*}\right)+\mathcal{O}(\varepsilon^{*2})\right], \label{vep-exp}
\end{align}
where $H^*=H(t(\tau^*))$ and $\varepsilon_i^*=\varepsilon_i(t(\tau^*))$.
Substituting them into $\mathcal{B}_{\text{LO}}$ gives
\begin{align}
    \mathcal{B}_{\text{LO}}(k_1,k_2,k_3)
    &=\frac{H^4}{32\varepsilon_1^2}\frac{1}{k_1^3k_2^3k_3^3}\big(\varepsilon_1K_1(k_1,k_2,k_3)+\varepsilon_2K_2(k_1,k_2,k_3)+\mathcal{O}(\tau)\big) \nonumber\\
    &=\frac{H^{*4}}{32\varepsilon_1^{*2}}\frac{1}{k_1^3k_2^3k_3^3}\Big(\varepsilon_1^*K_1(k_1,k_2,k_3)+\varepsilon_2^*K_2(k_1,k_2,k_3) \nonumber\\
    &\qquad+\ln\left(\frac{\tau}{\tau^*}\right)\big[\varepsilon_1^*(4\varepsilon_1^*+\varepsilon_2^*)K_1(k_1,k_2,k_3) \nonumber\\
    &\qquad\qquad\qquad\quad+\varepsilon_2^*(4\varepsilon_1^*+2\varepsilon_2^*-\varepsilon_3^*)K_2(k_1,k_2,k_3)\big]+\mathcal{O}(\varepsilon^{*3},\tau)\Big) \nonumber\\
\begin{split}
    &=\frac{H^{*4}}{32\varepsilon_1^{*2}}\frac{1}{k_1^3k_2^3k_3^3}\Big(\varepsilon_1^*K_1(k_1,k_2,k_3)+\varepsilon_2^*K_2(k_1,k_2,k_3) \\
    &\qquad-\ln\left(\frac{\tau}{\tau^*}\right)\big[\varepsilon_1^{*2}K_3(k_1,k_2,k_3)+\varepsilon_1^*\varepsilon_2^*K_4(k_1,k_2,k_3) \\
    &\qquad\qquad\qquad\quad+\varepsilon_2^{*2}K_5(k_1,k_2,k_3)+\varepsilon_2^*\varepsilon_3^*K_6(k_1,k_2,k_3)\big]+\mathcal{O}\big(\varepsilon^{*3},\tau\big)\Big).
\end{split}
\end{align}
For the NLO part $\mathcal{B}_{\text{NLO}}$, no further expansion is needed since $\mathscr{A}_{\text{NLO}}$ is already at second order:
\begin{align}
    \mathcal{B}_{\text{NLO}}(k_1,k_2,k_3)
    =\frac{H^4}{32\varepsilon_1^2}\frac{1}{k_1^3k_2^3k_3^3}\mathscr{A}_{\text{NLO}}
    =\frac{H^{*4}}{32\varepsilon_1^{*2}}\frac{1}{k_1^3k_2^3k_3^3}\Big(\mathscr{A}_{\text{NLO}}^*+\mathcal{O}\big(\varepsilon^{*3}\big)\Big),
\end{align}
where $\mathscr{A}_{\text{NLO}}^*$ is simply $\mathscr{A}_{\text{NLO}}$ with all $\varepsilon_i$ replaced by those evaluated at $\tau^*$ but $\ln(-k_{\text{t}}\tau)$ untouched. However, in their sum $\mathcal{B}_{\text{LO}}+\mathcal{B}_{\text{NLO}}$, the logarithms $\ln(-\tau)$ cancel out, and the total bispectrum $\mathcal{B}$ turns out to be equal to the expression \eqref{eq:BandAtau} with the replacements: $\varepsilon_i\rightarrow\varepsilon_i^*$, $H\rightarrow H^*$ and $\tau\rightarrow\tau^*$: 
\begin{align}
&\mathcal{B}(k_1,k_2,k_3)=\frac{H^{*4}}{32\varepsilon_1^{*2}}\frac{\mathscr{A}^*}{k_1^3k_2^3k_3^3}, 
\label{eq:BandA*1}\\
&\mathscr{A}^*=(\mathscr{A}_{\text{LO}}+\mathscr{A}_{\text{NLO}})_{\varepsilon_i\rightarrow\varepsilon_i^*,\,H\rightarrow H^*,\,\tau\rightarrow\tau^*}+\mathcal{O}(\varepsilon_i^{*3}). \label{eq:BandA*2}
\end{align}
Therefore, the bispectrum, which is defined at late time $\tau$, is independent of $\tau$ up to NLO as a consequence of the cancellation of $\ln(-\tau)$, meaning that the $\tau$-dependence of $H,\varepsilon_i$ in \eqref{eq:BandAtau} with \eqref{bispectrumtau} indeed cancels the logarithm $\ln(-\tau)$ in $\mathscr{A}_{\text{NLO}}$.
Furthermore, one can confirm that the bispectrum \eqref{eq:BandA*1} is independent of the choice of the pivot time $\tau^*$ by expanding $H^*,\varepsilon_i^*$ around yet another pivot time $\tau^\circ$, which gives \eqref{eq:BandA*1} with the replacement $(H^*,\varepsilon_i^*,\tau^*) \to (H^\circ,\varepsilon_i^\circ,\tau^\circ)$.

\subsection{Dominant contributions to non-gaussianity and its conserved definition}

We now discuss the dominant contributions to predictions of non-gaussianity in single field inflation models, encoded in the shape-dependent dimensionless parameter $f_\text{NL}$ defined by
\begin{align}\label{fNL}
    f_{\text{NL}}\equiv\frac{5}{12}\frac{\mathscr{A}^{*}}{k_1^3+k_2^3+k_3^3}\,.
\end{align}
At LO $f_\text{NL}$ is a function of ratios of momenta, while at NLO it acquires a logarithmic dependence, as seen in~\eqref{ANLO} with $\tau$ replaced by $\tau^*$. Note that $-\tau^* k$ is the physical momentum measured at time $\tau^*$ in units of the inflation scale $M_\text{I}\simeq H$: $-\tau^* k=k_\text{phys}/M_\text{I}=2\pi/(\lambda M_\text{I})$, with $\lambda$ some physical distance probed by CMB observations in the range $\sim 0.01-100$ Mpc, when the Universe was about 1000 times smaller than today. It follows that for momenta in this range, the dominant contribution to NLO relevant for observations comes from the logarithm $\ln(-\tau^* k_\text{t})$ in the first two lines of \eqref{ANLO}, as well as when a ratio of momenta becomes large near the squeezed limit that we study in the next section.

Using \eqref{ALO} and \eqref{ANLO} in \eqref{fNL}, one then finds~\cite{Antoniadis:2024abm}:
\begin{align}
\begin{split}
f_{\text{NL}}\simeq \frac{5}{12}&\bigg\{ \varepsilon_1\widetilde{K}_1+\varepsilon_2 +\left[-\varepsilon_1(\varepsilon_2+4\varepsilon_1)\widetilde{K}_1+\varepsilon_2(\varepsilon_3-2\varepsilon_2-4\varepsilon_1)\right]
  \ln(-\tau k_\text{t}) \\
&\qquad\qquad\quad\,
-\,(\varepsilon_2+2\varepsilon_1)^2\left(\ln{\frac{8k_1k_2k_3}{k_\text{t}^3}}\right)_\text{sq}
\bigg\}, \label{fNLdom}
\end{split} 
\end{align}
where we defined $\widetilde{K}_1(k_1,k_2,k_3) = K_1(k_1,k_2,k_3)/K_2(k_1,k_2,k_3)$\footnote{Note that this $\widetilde K_1$ is different from the definition in~\cite{Antoniadis:2024abm} by the numerical factor $5/12$.} and the term in the second line contributes only in the squeezed limit where one of the momenta is small, while we also dropped for the moment the star on $\tau$ for notational simplicity.

Note however that the definition \eqref{fNL} is not time independent since the proportionality factor between the bispectrum $\mathcal{B}$ and the 3-point amplitude $\mathscr{A}$ in \eqref{eq:BandAtau} acquires time dependence at the next order. Actually, using the lowest order expression of the power spectrum $\mathcal{P}_k^\text{LO}=H^2/(4k^3\varepsilon_1)$ and \eqref{eq:BandA*1}, one can rewrite \eqref{fNL} as
\begin{align}\label{fNLbis}
f_{\text{NL}}=\frac{5}{6}\frac{\mathcal{B}}{\mathcal{P}_{k_1}^\text{LO}\mathcal{P}_{k_2}^\text{LO}+\mathcal{P}_{k_2}^\text{LO}\mathcal{P}_{k_3}^\text{LO}+\mathcal{P}_{k_3}^\text{LO}\mathcal{P}_{k_1}^\text{LO}},
\end{align}
which is manifestly time dependent at NLO and beyond. A time independent definition would therefore be obtained by replacing in the denominator the LO power spectrum by its full conserved (time independent) expression~\cite{Burrage:2011hd}:
\begin{align}\label{fNLbis-hat}
\widehat f_{\text{NL}}=\frac{5}{6}\frac{\mathcal{B}(k_1,k_2,k_3)}{\mathcal{P}_{k_1}\mathcal{P}_{k_2}+\mathcal{P}_{k_2}\mathcal{P}_{k_3}+\mathcal{P}_{k_3}\mathcal{P}_{k_1}}.
\end{align}
Similarly to the bispectrum, the dominant contribution to the power spectrum up to NLO contains logarithmic corrections and reads\footnote{This is independent of $\tau$ as demonstrated in Appendix D of~\cite{Antoniadis:2024abm} and also in Section~\ref{subsec:different}.}~\cite{Antoniadis:2024abm}:
\begin{align}
\label{Pdom}
\mathcal{P}_k\simeq\frac{H^2}{4k^3\varepsilon_1}\left[1-(\varepsilon_2+2\varepsilon_1)\ln(-k\tau)\right].
\end{align}
It follows that $\widehat f_{\text{NL}}$ at NLO contains an extra term compared to the expression of $f_{\text{NL}}$ in \eqref{fNLdom}, given by
\begin{align}\label{fNLGNLO}
\begin{split}
\widehat f_{\text{NL,NLO}}&\simeq f_{\text{NL,NLO}} \\
&\quad+ (\varepsilon_2+2\varepsilon_1)f_\text{NL,LO}
\frac{k_3^3\ln(k_1k_2\tau^2)+k_2^3\ln(k_3k_1\tau^2)+k_1^3\ln(k_2k_3\tau^2)}{k_1^3+k_2^3+k_3^3},
\end{split}
\end{align}
where the symbol $\simeq$ denotes the dominant (logarithmic) contribution, as in \eqref{fNLdom} and \eqref{Pdom}.
In order to separate the two dominant contributions as in \eqref{fNLdom}, we multiply the arguments of the logarithms by $k_\text{t}^2/k_\text{t}^2$ and separate the universal part $\ln(k_\text{t}\tau)^2$ from those which are enhanced in the squeezed limits, to obtain:
\begin{align}\label{fNLGNLO2}
\frac{12}{5}\widehat f_{\text{NL,NLO}} &\simeq 
\left[-\varepsilon_1(\varepsilon_2+4\varepsilon_1)\widetilde{K}_1+\varepsilon_2(\varepsilon_3-2\varepsilon_2-4\varepsilon_1) 
+ 2(\varepsilon_2+2\varepsilon_1)(\varepsilon_1\widetilde{K}_1+\varepsilon_2)\right]
  \ln(-\tau k_\text{t})\nonumber\\
 &\quad -(\varepsilon_2+2\varepsilon_1)^2\left(\ln{\frac{8k_1k_2k_3}{k_\text{t}^3}}\right)_\text{sq}
 +(\varepsilon_2+2\varepsilon_1)\left((\varepsilon_1\widetilde{K}_1+\varepsilon_2)\ln{\frac{8k_1k_2k_3}{k_\text{t}^3}}\right)_\text{sq},
\end{align}
leading to the following expression for (the dominant contribution of) $\widehat f_\text{NL}$ up to NLO:
\begin{align}\label{fNLGNLO3}
\widehat f_\text{NL}\simeq \frac{5}{12}\left\{ \varepsilon_1\widetilde{K}_1+\varepsilon_2 +
\varepsilon_2(\varepsilon_1\widetilde{K}_1+\varepsilon_3)\ln(-\tau k_\text{t})\right\}.
\end{align}
It is now straightforward to show its time independence up to NLO by using the expansion formula \eqref{vep-exp}. Note that the squeezed limit divergence has been cancelled in the conserved version $\widehat f_\text{NL}$ of the parameter $f_\text{NL}$.

\section{Squeezed limit and consistency relation}

In this section, we compute the squeezed limit of the bispectrum up to NLO and the power spectrum up to next-to-next-to-leading order (N2LO) in the slow-roll parameters. We then verify that the consistency relation is still valid despite the presence of the logarithmically diverging factors which may invalidate the assumptions in the derivations of the consistency relation.

\subsection{Squeezed limit of the bispectrum and power spectrum}

Let us evaluate the squeezed limit $k_1\to0$ of the bispectrum up to NLO. We start with the bispectrum around the pivot scale $\tau^*$ \eqref{eq:BandA*1}. Since $k_1\sim0$, the other momenta have an almost equal size, $k\equiv k_2=k_3+\mathcal{O}(k_1)$. Using $K_1(k_1,k_2,k_3) \simeq 4k^3$ and $K_2(k_1,k_2,k_3) \simeq 2k^3$ in this limit, we obtain
\begin{align}
&\mathcal{B}(k_1,k_2,k_3)_{k_1\to0}=\frac{H^{*4}}{32k_1^3k^6\varepsilon_1^{*2}}\mathscr{A}^*_{k_1\rightarrow0}, 
\label{Bsqlimit} \\
\begin{split}
&\mathscr{A}^*_{k_1\rightarrow0}=2k^3\Bigg(2\varepsilon_1^*+\varepsilon_2^*+2\varepsilon_1^{*2}+5\varepsilon_1^*\varepsilon_2^*+2\varepsilon_2^{*2}-\varepsilon_2^*\varepsilon_3^*\\
&\qquad\qquad\qquad~ -\big(8\varepsilon_1^{*2}+6\varepsilon_1^*\varepsilon_2^*+2\varepsilon_2^{*2}-\varepsilon_2^*\varepsilon_3^*\big)\big[\gamma_E-1+\ln(-2k\tau^*)\big]\\
&\qquad\qquad\qquad~ -\big(2\varepsilon_1^*+\varepsilon_2^*\big)^2\ln\left(\frac{k_1}{k}\right)\Bigg)
\left[1+\mathcal{O}\left(\frac{k_1}{k},\varepsilon^{*3},\tau\right)\right]. \label{eq:limA}
    \end{split}
\end{align}

Let us proceed to the power spectrum.
As done for the bispectrum, it is convenient to introduce the amplitude $\mathscr{P}_k$ of the power spectrum as
\begin{align}
\label{PhatP}
\mathcal{P}_k=\frac{H^{*2}}{4k^3\varepsilon_1^*}\mathscr{P}^*_k, 
\end{align}
where we adopted the pivot scale $\tau^*$.
We have seen that the amplitude of the bispectrum $\mathscr{A}^*$ at NLO is of second order in $\varepsilon_i^*$.
In order to reproduce this from the power spectrum, we need $\mathscr{P}^*_k$ up to second order in $\varepsilon_i^*$, which is at the next-to-next-to-leading order (N2LO) for the power spectrum because $\mathscr{P}^*_k$ at LO is 1. In the late-time limit, the amplitude $\mathscr{P}^*_k$ up to N2LO with any pivot scale $\tau^*$ is given by~\cite{Gong:2001he,Schwarz:2001vv,Auclair:2022yxs,Bianchi:2024qyp,Antoniadis:2024abm}
\begin{align}\label{eq:hatPn2lo}
    \mathscr{P}^*_{k}=\mathcal{A}+\mathcal{B}\ln(-k\tau^*)+\mathcal{C}\big(\ln(-k\tau^*)\big)^2+\mathcal{O}\big(\varepsilon^{*3},\tau\big),
\end{align}
where the coefficients $\mathcal{A},\mathcal{B}$ and $\mathcal{C}$ are given by
\begin{align}
    \begin{split}&\mathcal{A}=1-2(\lambda+1)\varepsilon_1^*-\lambda\varepsilon_2^*+\frac{1}{2}\big(\pi^2+4\lambda^2+4\lambda-6\big)\varepsilon_1^{*2} \\
    &\quad+\frac{1}{12}\big(7\pi^2+12\lambda^2-12\lambda-72\big)\varepsilon_1^*\varepsilon_2^*+\frac{1}{8}\big(\pi^2+4\lambda^2-8\big)\varepsilon_2^{*2}+\frac{1}{24}\big(\pi^2-12\lambda^2\big)\varepsilon_2^*\varepsilon_3^*,\end{split}\\
    \begin{split}&\mathcal{B}=-2\varepsilon_1^*-\varepsilon_2^*+(4\lambda+2)\varepsilon_1^{*2}+(2\lambda-1)\varepsilon_1^*\varepsilon_2^*+\lambda\big(\varepsilon_2^{*2}-\varepsilon_2^*\varepsilon_3^*\big),\end{split}\\
    \begin{split}
        &\mathcal{C}=2\varepsilon_1^{*2}+\varepsilon_1^*\varepsilon_2^*+\frac{1}{2}\varepsilon_2^{*2}-\frac{1}{2}\varepsilon_2^*\varepsilon_3^*,\end{split}
\end{align}
with the constant $\lambda$ defined by $\lambda=\gamma_E+\ln(2)-2$.

\subsection{Consistency relation at NLO}
Finally, computing the right hand side of \eqref{eq:consistencyrelation} by using \eqref{PhatP} with \eqref{eq:hatPn2lo}, we obtain
\begin{align}
\begin{split}\label{eq:limARHS}
-\mathcal{P}_{k_1}\left(3+k_2\frac{d}{dk_2}\right)\mathcal{P}_{k_2}
&=\frac{H^{*4}}{16k_1^3k_2^3\varepsilon_1^{*2}}\Bigg(2\varepsilon_1^*+\varepsilon_2^*+2\varepsilon_1^{*2}+5\varepsilon_1^*\varepsilon_2^*+2\varepsilon_2^{*2}-\varepsilon_2^*\varepsilon_3^*\\
    &\qquad~-\big(8\varepsilon_1^{*2}+6\varepsilon_1^*\varepsilon_2^*+2\varepsilon_2^{*2}-\varepsilon_2^*\varepsilon_3^*\big)\big[\gamma_E-1+\ln(-2k_2\tau^*)\big]\\
    &\qquad~-\big(2\varepsilon_1^*+\varepsilon_2^*\big)^2\ln\left(\frac{k_1}{k_2}\right)\Bigg)+\mathcal{O}\big(\varepsilon^{*3},\tau\big),
    \end{split}
\end{align}
which, under $k=k_2=k_3+\mathcal{O}(k_1)$, is equal to the bispectrum in the squeezed limit \eqref{Bsqlimit} with \eqref{eq:limA}, thereby verifying the consistency relation at NLO. Note that this has been  verified for any choice of the pivot scale $\tau^*$.

\subsection{Consistency relation with different horizon exit times}
\label{subsec:different}
Actually, one could argue that the long wavelength mode $k_1$ exits the horizon at an earlier time $\tau^\sharp$ than the long wavelength modes $k_2$ and $k_3$ which cross the horizon at $\tau^\flat$. Namely we have $\tau^\sharp<\tau^\flat$ and one could therefore guess that the consistency relation takes the form 
\begin{align}\label{eq:consistencydifferenttimes}
    &\mathcal{B}(k_1,k_2,k_3)_{k_1\rightarrow0}=-\mathcal{P}_{k_1}^\sharp
\left(3+k_2\frac{\partial}{\partial k_2}\right)\mathcal{P}_{k_2}^\flat\left[1+\mathcal{O}\left(\frac{k_1}{k_2}\right)\right],
\end{align}
as it was originally guessed \cite{Maldacena:2002vr}. Here $\mathcal{P}_{k_1}^\sharp$ and $\mathcal{P}_{k_1}^\flat$ are \eqref{PhatP} where $(H^*,\varepsilon_i^*,\tau^*)$ is replaced by $(H^\sharp,\varepsilon_i^\sharp,\tau^\sharp)$ and $(H^\flat,\varepsilon_i^\flat,\tau^\flat)$, respectively.
However, we want to argue that this distinction does not modify the result. Indeed, the power spectrum can be computed explicitly at NLO in terms of the Hubble and slow-roll parameters at $\tau$ as
\begin{align}
    \mathcal{P}_k=\frac{H^2}{4k^3\varepsilon_1}\left[1+\varepsilon_2-(2\varepsilon_1+\varepsilon_2)\left(\gamma_E-1+\ln(-2k\tau)\right)\right]+\mathcal{O}\big(\varepsilon,\tau\big),
\end{align}
and we can explicitly check that, as the bispectrum, it takes exactly the same form when we expand the parameters around a pivot scale $\tau^*$ in the late-time limit:
\begin{align}
    \mathcal{P}_k=\frac{H^{*2}}{4k^3\varepsilon_1^*}\left[1+\varepsilon_2^*-(2\varepsilon_1^*+\varepsilon_2^*)\left(\gamma_E-1+\ln(-2k\tau^*)\right)\right]+\mathcal{O}\big(\varepsilon^*,\tau\big),
\end{align}
which is indeed \eqref{eq:hatPn2lo} up to NLO. In general, the power spectrum is expected to take the following form:
\begin{align}
    \mathcal{P}_{k}=\frac{H^{*2}}{4k^3\varepsilon_1^*}\sum\limits_{n=0}^{\infty}\mathcal{A}_n\big(\ln(-k\tau^*)\big)^n+\mathcal{O}\big(\varepsilon^{*2},\tau\big)
\end{align}
with $\mathcal{A}_n$ some polynomials of the $\varepsilon_i^*$, when computed around an arbitrary pivot scale\footnote{This was explicitly checked at N3LO in \cite{Auclair:2022yxs,Bianchi:2024qyp,Antoniadis:2024abm}.}. Therefore, the dependence on $\tau$, the time at which the 2-point function is actually computed, is subleading in the late-time limit. Consequently \eqref{eq:consistencydifferenttimes} can be put in the form
\begin{align}
    &\mathcal{B}(k_1,k_2,k_3)_{k_1\rightarrow0}=-\mathcal{P}_{k_1}^*
\left(3+k_2\frac{\partial}{\partial k_2}\right)\mathcal{P}_{k_2}^*\left[1+\mathcal{O}\left(\frac{k_1}{k_2},\tau^\sharp,\tau^\flat\right)\right],
\end{align}
and is therefore equivalent to \eqref{eq:consistencyrelation}.

\section{Conclusion}
We have seen that despite the infrared diverging contributions at large physical distances
and in the squeezed limits, the consistency relation \eqref{eq:consistencyrelation} relating the squeezed limit of the bispectrum to the power spectrum holds at late-time, at NLO in slow-roll parameters for single field inflation models, where the squeezed limit of the amplitude of the bispectrum is given by \eqref{eq:limA} with any pivot time $\tau^*$ that covers the time $\tau$ of the correlator itself. Note that the consistency relation has been verified even in the presence of the small but finite cutoff $k_1$ in the form $\ln(k_1/k)$ as demonstrated in \eqref{eq:limA}. In the course of the verification, we have shown that the bispectrum at late time \eqref{eq:BandAtau} computed in~\cite{Antoniadis:2024abm} is actually time independent due to the cancellation of the time dependence of $H,\varepsilon_i$ and the logarithm $\ln(-\tau)$ in \eqref{eq:BandAtau}, and also that it is independent of the pivot time at which $H,\varepsilon_i$ are evaluated. We also have shown that the power spectrum also has exactly the same property: it is independent of the time at which the 2pt function is defined and of the choice of the pivot time at which the Hubble and slow-roll parameters are defined.

Let us recapitulate it. At leading order, although the bispectrum is divergent in this limit due to the overall $k_1^{-3}$ factor, the amplitude $\mathscr{A}^*$ defined by 
\begin{align}
    \mathcal{B}(k_1,k_2,k_3)=\frac{H^{*4}}{32\varepsilon_1^{*2}}\frac{\mathscr{A}^{*}}{k_1^3k_2^3k_3^3}
\end{align}
is finite in this limit at LO in $\varepsilon^{*}$. However, this is not true at higher orders. Indeed, we have found at NLO two logarithms, $\ln(-k_\text{t}\tau)$ that diverges in the large physical distance limit and $\ln(k_1/k)$ that diverges in the squeezed limit~\cite{Antoniadis:2024abm}. Therefore, the usual $f_{\text{NL}}$ parameter defined by~\eqref{fNL}, giving, with $k=k_2=k_3+\mathcal{O}(k_1)$,
\begin{align}
\begin{split}
    f_{\text{NL},k_1\rightarrow0}&=\frac{5}{12}\Bigg(2\varepsilon_1^{*}+\varepsilon_2^{*}+2\varepsilon_1^{*2}+5\varepsilon_1^{*}\varepsilon_2^{*}+2\varepsilon_2^{*2}-\varepsilon_2^{*}\varepsilon_3^{*}\\
    &\qquad\qquad-\big(8\varepsilon_1^{*2}+6\varepsilon_1^{*}\varepsilon_2^{*}+2\varepsilon_2^{*2}-\varepsilon_2^{*}\varepsilon_3^{*}\big)\big[\gamma_E-1+\ln(-2k\tau^{*})\big]\\
    &\qquad\qquad-\big(2\varepsilon_1^{*}+\varepsilon_2^{*}\big)^2\ln\left(\frac{k_1}{k}\right)\Bigg)+\mathcal{O}\left(\frac{k_1}{k},\varepsilon^{*3},\tau\right),
\end{split}
\end{align}
diverges as $k_1$ goes to zero. Despite these logarithms, we have shown that the consistency relation \eqref{eq:consistencyrelation} is still valid in the vicinity of this limit and at large but finite time.

This $f_{\text{NL}}$ depends on $\tau$ because it was defined by dividing $\mathcal{B}$ up to NLO by the power spectrum at LO. We have then introduced a modified version $\widehat f_{\text{NL}}$ by dividing $\mathcal{B}$ up to NLO by the power spectrum also up to NLO, and demonstrated that $\widehat f_{\text{NL}}$ is now time-independent up to NLO and also the squeezed limit divergence is removed in $\widehat f_{\text{NL}}$.

It would be of course interesting to investigate whether the consistency relation remains valid at  higher orders in the slow-roll parameters, generalising the existing proofs in the presence of a finite time cutoff and in the vicinity of the squeezed limit of the bispectrum. This amounts to showing that the limit of the ratio $\mathcal{B}_{k_1,k_2,k_3}/\mathcal{P}_{k_1}$ as $k_1\to 0$ exists to all orders at late time.

\section*{Acknowledgements}

I.A. and H.I. are supported by the Second Century Fund (C2F), Chulalongkorn University. A.C. has been supported by Thailand NSRF via PMU-B, grant number B39G670016. We are grateful to Spyros Sypsas, Shinji Tsujikawa and Vicharit Yingcharoenrat for fruitful discussions.


\providecommand{\href}[2]{#2}\begingroup\raggedright\endgroup

\end{document}